\documentclass[12pt]{article}
\usepackage{graphicx}
\usepackage{amsmath}

\begin{document}
\date{}
\title{Neutrino masses and mixings with an S$_{3}$ family permutation symmetry}
\author{{\small Francesco Caravaglios and Stefano Morisi} \\
{\small Dipartimento di Fisica, Universit\`{a} di Milano, Via Celoria 16,
 I-20133 Milano, Italy}
\\ {\small and}\\{\small INFN\ sezione di Milano}}
\maketitle

\begin{abstract}
Large neutrino mixing angles suggest that the Yukawa sector is invariant
under permutations of the fermion families. This S$_{3}$ permutation
symmetry is broken at a large energy scale but much below the unification \
scale. Assuming that the lepton mass matrix is approximately diagonal, all
neutrino mixing angles naturally come from the breaking of S$_{3}\rightarrow 
$S$_{2}.$ In the neutrino sector, S$_{2}$ remains (approximately) unbroken.
As a consequence, we have a large atmospheric neutrino angle and U$_{e3}=0$.
The S$_{3}$ symmetry at the unification scale can also explain the large
solar mixing angle. We give an explicit expression of the solar mixing angle in
 terms of the left-handed neutrino masses.\\
 We observe that this family permutation symmetry comes
very naturally from a quantized theory of functionals\cite{carav1}, that is
an extension of quantum field theory.
\end{abstract}

\section{Introduction}

In the Standard Model (SM) Lagrangian, the breaking of the SU(2)$\times $%
U(1) gauge symmetry is due to a Higgs \ scalar doublet. This is the most
realistic possibility since triplets or larger representations give
unrealistic masses and mixings angles in the electroweak gauge boson sector.
In order to give mass to the SM fermions, we introduce Yukawa interactions
between fermions and the Higgs doublet.

Neutrinos are very light with respect to all other fermions. The reason is
probably that the right-handed neutrino is SU(2)$\times $U(1) singlet. It
takes a Majorana mass that does not \ break the electroweak gauge symmetry,
and so its mass is expected to be of the order of the Grand Unification
scale. The left-handed neutrino has a Majorana mass that is proportional to
the inverse of the right-handed neutrino mass. Through this seesaw mechanism
we understand why neutrinos are so light.

Also neutrino mixings appear very different from quark mixings. The $V_{%
\text{ckm}}$ matrix is close to the identity matrix, mixing angles are not
large. The mixing between the heaviest quarks with the lightest is very
small. On the contrary, recent neutrino experiments[2-6] have shown that 
mixings angles between neutrinos with
different lepton flavour are very large, approximately maximal. A good
candidate for the MNS matrix, (the analogue of the CKM\ matrix) is 
\begin{equation}
O_{\text{MNS}}=\left( 
\begin{tabular}{lll}
$\frac{-2}{\sqrt{6}}$ & $\frac{1}{\sqrt{3}}$ & $0$ \\ 
$\frac{1}{\sqrt{6}}$ & $\frac{1}{\sqrt{3}}$ & $\frac{-1}{\sqrt{2}}$ \\ 
$\frac{1}{\sqrt{6}}$ & $\frac{1}{\sqrt{3}}$ & $\frac{1}{\sqrt{2}}$%
\end{tabular}
\right) .  \label{1}
\end{equation}
Since we measure $\sin ^{2}\theta _{\text{sol}}$ \ and $\sin ^{2}\theta _{%
\text{atm}}$ the choice of signs in the matrix elements (\ref{1}) is not
unique. The choice in (\ref{1}) corresponds to neutrino mass eigenstates
that are also eigenstates of the permutation symmetry $\nu _{\mu
}\leftrightarrow \nu_{\tau }$. This S$_{2}$ symmetry directly implies $%
U_{e3}=0 $ and a maximal atmospheric neutrino mixing angle (see the third
column in (\ref{1})). The importance of discrete symmetries in the
description of neutrino masses and mixings has been observed by several
authors[7-20].
In this paper we will show that if we embed S$_{2}$ into a S$_{3}$ symmetry,
that is the permutation symmetry of the three fermion families (including
the three right-handed neutrinos), and if we add just one \ (left-handed)\
Weyl fermion\footnote{%
However this additional fermion $\chi _{L}$ is not necessary. It helps in
the explanation of the large splitting between S$_{3}$ singlet and doublet
right-handed neutrinos.} $\chi _{L}$, that is both a SM\ singlet and a S$%
_{3} $ singlet, we can predict \ the MNS\ matrix in eq.(\ref{1}). This S$%
_{3}$ family symmetry arises in a very natural way from a quantized theory
of functionals\cite{carav1}, that is an extension of quantum field theory.
We will discuss a simple model with three right-handed neutrinos, where the S%
$_{3}$ symmetry is spontaneously broken into S$_{2}$. The right-handed
neutrino can get a Majorana mass only when the additional U(1) (included in
SO(10)) is broken. Before the breaking of the S$_{3}$ symmetry, $\chi _{L}$
only mixes with the S$_{3}$ singlet right-handed neutrino. This introduces a
splitting between the S$_{3}$ singlet and doublet in the $\nu_{R}$ sector. The
solar angle $\sin \theta _{\text{sol}}=1/\sqrt{3}$ arises when this
splitting is large, while the breaking \ S$_{3}\rightarrow $S$_{2}$ implies U%
$_{e3}=0$ and $\sin \theta _{\text{atm}}=1/\sqrt{2}.$ In the next section we
will discuss a generic S$_{2}$ symmetric matrix, and in the third section we
will discuss the S$_{3}$ symmetric case.

\section{\protect\bigskip S$_{2}$ symmetry and neutrino mass and mixings}

First, we discuss the most general $S_{2}$ symmetric real mass matrix. We
require that the mass matrix is invariant when we exchange the 2nd with the
3rd row (and column), 
\begin{equation}
P\,\ M_{\nu}^{R}\,P=M_{\nu}^{R}  \label{bis}
\end{equation}
where $P$ is the S$_{2}$ \ permutation matrix 
\begin{equation*}
P=\left( 
\begin{array}{ccc}
1 & 0 & 0 \\ 
0 & 0 & 1 \\ 
0 & 1 & 0
\end{array}
\right) .
\end{equation*}
The matrix equation (\ref{bis}) is satisfied if and only if \ 
\begin{equation}
M_{\nu}^{R}=\left( 
\begin{array}{ccc}
a & d & d \\ 
d & b & c \\ 
d & c & b
\end{array}
\right)  \label{9}
\end{equation}
where $a,b,c,d$ are four real parameters. Since all parameters are real, we
need just an orthogonal matrix $O$ to diagonalize the symmetric matrix $%
M_{\nu}^{R},$ \ (\ref{9}). We already know that the vector $v_{3}=(0,-1/\sqrt{2%
},1/\sqrt{2})$ is an eigenvector, since it is the only state with odd $S_{2}$
parity: $Pv_{3}=-v_{3}$. So, two of the three angles of the orthogonal
matrix $O$ are fixed by the $S_{2}$ symmetry, while the third angle ( the
solar neutrino mixing angle) that mixes the two states orthogonal to $\nu
_{3}$ is free. Any matrix in (\ref{9}) can be diagonalized by the orthogonal
matrix

\begin{equation}
O=\left( 
\begin{array}{ccc}
-\cos \theta & \sin \theta & 0 \\ 
\frac{1}{\sqrt{2}}\sin \theta & \frac{1\ }{\sqrt{2}}\cos \theta & -\frac{1}{%
\sqrt{2}} \\ 
\frac{1}{\sqrt{2}}\sin \theta & \frac{1}{\sqrt{2}}\cos \theta & \frac{1}{%
\sqrt{2}}
\end{array}
\right)  \label{5}
\end{equation}
and the angle $\sin \theta $ is obtained by imposing that the matrix $m_{\nu
}=O^{t}M_{\nu}^{R}O$ \ must be diagonal. 
\begin{equation}
m_{\nu}=O^{t}\ \left( 
\begin{array}{ccc}
a & d & d \\ 
d & b & c \\ 
d & c & b
\end{array}
\right) O  \label{10}
\end{equation}
or more explicitly 
\begin{equation}
m_{\nu}=\ \left( 
\begin{array}{ccc}
z+x\cos 2\theta -y\sin 2\theta & -x\sin 2\theta -y\cos 2\theta & 0 \\ 
-x\sin 2\theta -y\cos 2\theta & z-x\cos 2\theta +y\sin 2\theta & 0 \\ 
0 & 0 & b-c
\end{array}
\right)   \label{11}
\end{equation}
where 
\begin{equation}
\begin{array}{cc}
x= & \frac{1}{2}\,(a-b-c) \\ 
y= & \sqrt{2}\,d \\ 
z= & \frac{1}{2}\,(a+b+c).
\end{array}
\label{11b}
\end{equation}
We have to set to zero the off diagonal entries, and from this condition we
derive the solar mixing angle 
\begin{equation}
\tan 2\theta =-\frac{y}{x}  \label{12}
\end{equation}
or 
\begin{equation}
\sin 2\theta =\frac{y}{\sqrt{x^2+y^{2}}}  \label{13}
\end{equation}
If we put this value of $\theta $ in the matrix (\ref{5}), the matrix (\ref
{11}) is diagonal and the elements in the diagonal give the three masses 
\begin{equation}
\begin{array}{cc}
m_{1}^{R}= & z-\sqrt{x^{2}+y^{2}} \\ 
m_{2}^{R}= & z+\sqrt{x^{2}+y^{2}} \\ 
m_{3}^{R}= & b-c.
\end{array}  \label{12b}
\end{equation}

\section{S$_{3}$ symmetry breaking and neutrino masses and mixings}

We consider a S$_{3}$ symmetric model with three left-handed neutrinos $\nu
_{L}^{i}$ ($i$ is the family index) and three right-handed neutrinos $\nu
_{R}^{i}$. We also add a left-handed Weyl fermion $\chi _{L}$, that is a
Standard Model singlet (SU(3)$\times $SU(2)$\times $U(1) singlet). It does
not carry the family index, i.e. it is a S$_{3}$ singlet. The scalar sector
includes two standard model singlets for each family, $\phi ^{i}$ and $%
\varphi ^{i}$, plus the common SU(2) scalar doublet $H$.   We assume that
the lepton mass matrix is approximately diagonal, while for the neutrinos we
consider a Yukawa interaction of the form\footnote{%
The scalar field $\varphi ^{i}$ does not break S$_{3}$: ($\left\langle
\varphi ^{1}\right\rangle =\left\langle \varphi ^{2}\right\rangle
=\left\langle \varphi ^{3}\right\rangle $)} 
\begin{equation}
L_{\text{yuk}}^{a}=g\,\sum_{i}\bar{\nu}_{L}^{i}\,\nu _{R}^{i}\,H+\lambda
_{1}\sum_{i}\,\bar{\nu}_{R}^{c\,\ i}\,\nu _{R}^{i}\,\,\phi ^{i}+\lambda
_{2}\,\sum_{i}\bar{\chi}_{L}^{\,}\,\nu _{R}^{i}\,\,\varphi ^{i}+M\,\bar{\chi}%
_{L}^{\,}\,\chi _{L}^{c}+{\text{h.c.}}
\end{equation}
we have added the S$_{3}\times $SU(3)$\times $SU(2)$\times $U(1) singlet, $%
\,\chi _{L}^{c}$, because we want an explicitly renormalizable Lagrangian
and we want to explain the splitting among the three right-handed neutrinos,
but such a singlet is not necessary since one could add such a splitting
directly to the Lagrangian (without breaking S$_3$) 
\begin{equation}
L_{\text{yuk}}^{b}=g\sum_{i}\,\bar{\nu}_{L}^{i}\,\nu
_{R}^{i}\,H+\sum_{i}\lambda _{1}\,\bar{\nu}_{R}^{c\,\ i}\,\nu
_{R}^{i}\,\,\phi ^{i}+\lambda _{3}\,\,\,\sum_{i,j}\bar{\nu}_{R}^{c\,\
i}\,\nu _{R}^{j}\,\,\psi+{\text{h.c.}}   \label{4}
\end{equation}
$L_{\text{yuk}}^{b}$ is simpler, while $L_{\text{yuk}}^{a}$ and 
the S$_3$ singlet $\chi_L$ can probably
 explain   the large splitting among 
the right-handed neutrinos in terms of
additional symmetries of the Lagrangian. In any case, there are two
important ingredients that appear to be necessary in realistic examples:
there must be an S$_{3}$ symmetry, at the energy scale where the
 heaviest right-handed neutrino acquires a mass\footnote{%
That is probably the scale at which the extra U(1) included in SO(10) is
broken.}, then S$_{3}$ must break at a much lower scale. The breaking S$%
_{3}\rightarrow $S$_{2}$ occurs at a much smaller scale than the grand
unification scale.

To start we take the Lagrangian (\ref{4}). $\psi $ and $\phi ^{i}$ are two
scalar fields, with $\left\langle \psi \right\rangle >>$ $\left\langle \phi
^{i}\right\rangle $. $H$ is the Higgs SU(2) doublet, it does not carry any
family index\footnote{%
This is the most realistic situation, taking into account constraints coming
from electroweak precision tests of the Standard Model.}. To simplify the
notation we set $\lambda _{3}=1$, since it is an irrelevant scale factor.
The mass matrix for the right-handed neutrinos becomes 
\begin{equation}
M_{\nu }^{R}=\left\langle \psi \right\rangle \left( 
\begin{tabular}{ccc}
$\frac{\left\langle \phi ^{1}\right\rangle }{\left\langle \psi \right\rangle 
}+1$ & $1$ & $1$ \\ 
$1$ & $\ \frac{\left\langle \phi ^{2}\right\rangle }{\left\langle \psi
\right\rangle }+1$ & $1$ \\ 
$1$ & $1$ & $\ \frac{\left\langle \phi ^{3}\right\rangle }{\left\langle \psi
\right\rangle }+1$%
\end{tabular}
\right).  \label{massar}
\end{equation}
Now we assume $\left\langle \phi ^{2}\right\rangle =\left\langle \phi
^{3}\right\rangle $ such that the S$_{2}$ exchange symmetry remains unbroken,
and $\left\langle \psi \right\rangle >>$ $\left\langle \phi
^{i}\right\rangle $. The Dirac mass matrix $g\,\bar{\nu}_{L}^{i}\,\nu
_{R}^{i}\,H$ that mixes left-handed neutrinos with their right-handed
component is proportional to the identity matrix. Then, due to the seesaw,
the left-handed neutrinos get a Majorana mass matrix that is proportional
just to the inverse of the matrix $M_{\nu }^{R}$. Namely, 
\begin{equation}
M_{\nu }^{L}=g^{2}\,\left\langle H\right\rangle ^{2}\,\left( M_{\nu
}^{R}\right) ^{-1}.  \label{nl}
\end{equation}
The same orthogonal matrix will diagonalize both $\,M_{\nu }^{R}$ and $%
M_{\nu }^{L}$. This means that is sufficient to diagonalize$\,M_{\nu }^{R}$.
\ If $\left\langle \phi ^{2}\right\rangle =\left\langle \phi
^{3}\right\rangle ,$ the matrix (\ref{massar}) is S$_{2}$ symmetric, and it
is equivalent to the (\ref{9}) with $c=d$. From eqs.(\ref{11b},\ref{12},\ref
{12b}) we \ \ find a relation that constrains the solar mixing angle to be a
function of the right-handed neutrino masses ($m_{1}^{R},m_{2}^{R}$ and $%
m_{3}^{R}$)

\begin{equation*}
\sqrt{2}\sin 2\,\theta _{\text{sol}}-\cos 2\,\theta _{\text{sol}}=\frac{%
m_{1}^{R}+m_{2}^{R}-2\,m_{3}^{R}}{m_{2}^{R}-m_{1}^{R}}.
\end{equation*}

We can also express the solar mixing angle in terms of the left-handed
neutrino masses\footnote{%
The left-handed neutrino mass eigenvalues are proportional to the inverse of
the right-handed neutrino masses.}

\begin{equation}
\sqrt{2}\sin 2\,\theta _{\text{sol}}-\cos 2\,\theta _{\text{sol}}=\frac{%
m_{2}^{L}m_{3}^{L}+m_{1}^{L}m_{3}^{L}-2\,m_{1}^{L}m_{2}^{L}}{%
m_{1}^{L}m_{3}^{L}-m_{2}^{L}m_{3}^{L}}=R
\end{equation}
or 
\begin{equation*}
\theta _{\text{sol}}=\frac{1}{2}\arcsin \left( \frac{R}{\sqrt{3}}\right) +%
\frac{1}{2}\arcsin \left( \frac{1}{\sqrt{3}}\right).
\end{equation*}
We remind that $m_{3}^{L}$ is the
mass of left-handed neutrino corresponding to the S$_{2}$ parity odd state.
Note that in the limit $m_{2}^{R}\rightarrow \infty $, we have $%
m_{2}^{L}\sim 1/m_{2}^{R}\rightarrow 0$, 
 and $\sin \theta _{\text{sol}}=1/\sqrt{3}$.  
This limit is achieved when the v.e.v. $\left\langle\psi\right\rangle$ is
 much larger than $\left\langle \phi^i \right\rangle$. The first v.e.v
$\left\langle\psi\right\rangle$  breaks SO(10) and leave S$_3$ unbroken,
while the second one  breaks S$_3$.
 This is the S$%
_{3}$ approximately symmetric limit, when the S$_{3}$ breaking scale is very
small compared to the U(1)\ breaking scale and unification scale.

\section{Conclusions}

We have studied the breaking of an S$_{3}$ symmetry, that is the permutation
of the three fermion families. This breaking naturally explains the large
mixing angles among neutrinos of different flavours. We have concluded that
while the unbroken S$_{2}$ permutation of the $\mu $ neutrino with the $\tau 
$ neutrino predicts both the atmospheric mixing angle and U$_{e3}=0$, the S$%
_{3}$ symmetry below the scale of the U(1)$\in $SO(10) breaking gives a
simple mechanism to explain the solar neutrino mixing angle. A very simple
Yukawa sector, S$_{3}$ symmetric like 
\begin{equation*}
L_{\text{yuk}}^{b}=g\,\sum_{i}\,\,\bar{\nu}_{L}^{i}\,\nu
_{R}^{i}\,\left\langle H\right\rangle +\lambda _{1}\,\sum_{i}\,\,\bar{\nu}%
_{R}^{c\,\ i}\,\nu _{R}^{i}\,\,\left\langle \phi ^{i}\right\rangle +\lambda
_{3}\,\sum_{i,j}\,\,\bar{\nu}_{R}^{c\,\ i}\,\nu _{R}^{j}\,\,\left\langle
\psi \right\rangle \,
\end{equation*}
after the breaking of  S$_3$ into S$_2$ automatically gives 
\begin{equation*}
\begin{tabular}{l}
U$_{e3}=0$ \\ 
$\sin \,\theta _{\text{atm}}=1/\sqrt{2}$ \\ 
$\sqrt{2}\sin 2\,\theta _{\text{sol}}-\cos 2\,\theta _{\text{sol}}=
(m_{2}^{L}m_{3}^{L}+m_{1}^{L}m_{3}^{L}-2%
\,m_{1}^{L}m_{2}^{L})/(m_{1}^{L}m_{3}^{L}-m_{2}^{L}m_{3}^{L}).$%
\end{tabular}
\end{equation*}
The last condition also gives $\sin \,\theta _{\text{sol}}=1/\sqrt{3}$ in
the limit $m_{2}^{L}=0.$
This limit suggests that the breaking of S$_3$ occurs at an energy scale much below the unification breaking scale.
Finally, we observe that the family permutation symmetry derives from a
quantized theory of functionals\cite{carav1}, that is an extension of
quantum field theory.

\end{document}